\documentclass[prd,amsmath,amssymb,nofootinbib,preprintnumbers]{revtex4}

\usepackage{graphicx}
\usepackage{dcolumn}
\usepackage{bm}
\usepackage{amsmath}
\usepackage{amsfonts}

\def\ls{\mathrel{\lower4pt\vbox{\lineskip=0pt\baselineskip=0pt
           \hbox{$<$}\hbox{$\sim$}}}}
\def\gs{\mathrel{\lower4pt\vbox{\lineskip=0pt\baselineskip=0pt
           \hbox{$>$}\hbox{$\sim$}}}}
\def\drawbox#1#2{\hrule height#2pt

\hbox{\vrule width#2pt height#1pt \kern#1pt
              \vrule width#2pt}
              \hrule height#2pt}

\def\Asym#1#2{\vcenter{\vbox{\drawbox{#1}{#2}
              \kern-#2pt       
              \drawbox{#1}{#2}}}}

\newcommand{\beq}{\begin{equation}}
\newcommand{\eeq}{\end{equation}}

\begin{document}

%
\title{Inflation and The Minimal Supersymmetric Standard Model}

\author{Rouzbeh Allahverdi}

\affiliation{Department of Physics \& Astronomy, University of New Mexico, Albuquerque, NM 87131, USA
}



\begin{abstract}
There is strong evidence from cosmological data that the universe underwent an epoch of superluminal expansion called inflation. A satisfactory embedding of inflation in fundamental physics has been an outstanding problem at the interface of cosmology and high energy physics.
We show how inflation can be realized within the Minimal Supersymmetric Standard Model (MSSM).
The inflaton candidates are two specific combinations of supersymmetric partners of quarks and leptons. MSSM inflation occurs at a low scale
and generates perturbations in the range experimentally allowed
by the latest data from Wilkinson Microwave Anisotropy Probe (WMAP).
The parameter space for inflation is compatible with supersymmetric dark matter, and the Large Hadron Collider (LHC) is capable of discovering the inflaton candidates in the allowed regions of parameter space.
\end{abstract}

\maketitle

\section{Introduction}

Inflation is the dominant paradigm of the early universe cosmology. It solves the flatness and isotropy problems of the hot big bang cosmology and generates the seeds for structure formation~\cite{Linde}. To date, experiments confirm simplest predictions of inflation: a flat universe and nearly scale-invariant adiabatic fluctuations with a gaussian spectrum, which are imprinted in temperature anisotropy of the Cosmic Microwave Background (CMB)~\cite{WMAP5}. In spite of its impressive success, a satisfactory explanation of the microscopic origin of inflation has been lacking over the past 27 years. In almost all proposed models~\cite{LR} the inflaton is added as an absolute gauge singlet which has no natural place in particle physics~\footnote{For the only exceptions known to the author, see Refs.~\cite{MSYY,LS,KMT}.}. The inflaton mass and its couplings to other fields are not generally tied to any fundamental theory and instead are set by hand in order to match observations. In particular, since the inflaton couplings to the Standard Model (SM) fields are arbitrary, it is not clear how most of the inflaton energy eventually goes into the observable sector, as required by the primordial abundance of light elements made during Big Bang Nulceosynthesis (BBN)~\cite{BBN}.

The Minimal Supersymmetric Standard Model (MSSM) is a well motivated extension of the SM~\cite{Nilles}. It introduces scalar partners for the quarks and leptons (called squarks and sleptons respectively), and fermionic partners for the gauge and Higgs fields (called gauginos and Higgsinos respectively). The supersymmetric partners have the same quantum numbers as ordinary fields, and their mass is supposed to be in the $\sim 100~{\rm GeV}-1$ TeV range in order to address the hierarchy problem of the SM. There exist a large number of flat directions in the field space, consisting of the squark and slepton fields, along which the scalar potential identically vanishes in the limit of exact supersymmetry~\cite{GKM}. These flat directions have important cosmological consequences~\cite{EM,DK}. In particular, it has been recently shown that two specific flat directions can lead to a successful inflation~\cite{AEGM,AEGJM}. This is the first example of the inflaton finding a natural place in a well motivated extension of the SM. In particular, since the inflaton is related to the squark and slepton fields, the inflation sector can be probed in colliders, most notably the Large Hadron Collider (LHC).

The aim of this article is to present a brief review of MSSM inflation and its various achievements. In Section 2 we show how inflation can happen along MSSM flat directions. We then discuss properties of MSSM inflation and its predictions in light of the 5-year Wilkinson Microwave Anisotropy (WMAP) data in Section 3. Next we turn to the compatibility between MSSM inflation and supersymmetric dark matter and present a combined analysis of parameter space in the case of minimal supergravity in Section 4. In Section 5 we briefly discuss the reheating of the universe after MSSM inflation. Section 6 contains conclusions and some discussions.

\section{Inflation in MSSM}

Flat directions in the scalar potential of MSSM are classified by gauge-invariant monomials made of the squark, slepton and Higgs fields~\cite{GKM}. In the limit of unbroken supersymmetry, the potential identically vanishes along these directions. Supersymmetry breaking terms and non-renormalizable superpotential terms lift the flat directions~\cite{DRT1,DRT2}. Denoting the flat direction by $\Phi$, the superpotential terms have the form
\begin{equation} \label{non}
W \supset \lambda {\Phi^n \over n M^{n-3}},
\end{equation}
where $n > 3$ and $M$ is the scale of the new physics that induces these terms. Planck-suppressed non-renormalizable terms are generically expected to be induced by quantum gravity or string theory, for which $M = M_{\rm P} \equiv 2.4 \times 10^{18}$ GeV and $\lambda_n \sim {\cal O}(1)$. This is the case that we consider henceforth. Within the MSSM all flat directions are lifted by non-renormalizable operators with $4 \leq n \leq 9$~\cite{GKM}.

The scalar potential along the flat direction reads~\cite{AEGM}
\begin{equation} \label{flatpot}
V = {1\over2} m^2_\phi\,\phi^2 + A\cos(n \theta  + \theta_A)
{\lambda \phi^n \over n\,M^{n-3}_{\rm P}} + \lambda^2
{{\phi}^{2(n-1)} \over M^{2(n-3)}_{\rm P}}\, ,
\end{equation}
where the first and last terms on the right-hand side are the soft mass term and the $A$-term respectively. Here $\phi$ and $\theta$ denote the radial and the angular coordinates of the complex scalar field
$\Phi=\phi\,\exp[i\theta]$, while $\theta_A$ is the phase of the $A$-term (thus $A$ is a positive quantity with dimension of mass). For weak scale supersymmetry and gravity mediation of supersymmetry breaking to the observable sector we have $m_{\phi} \sim A \sim {\cal O}({\rm TeV})$. After minimizing the potential along the angular direction we find
\begin{equation} \label{scpot}
V(\phi) = {1\over2} m^2_\phi\,\phi^2 - A {\lambda \phi^n \over n\,M^{n-3}_{\rm P}} + \lambda^2 {{\phi}^{2(n-1)} \over M^{2(n-3)}_{\rm P}} .
\end{equation}
If
\begin{equation} \label{dev}
{A^2 \over 8 (n-1) m^2_{\phi}} \equiv 1 + \Big({n-2 \over 2}\Big)^2
\alpha^2\, ,
\end{equation}
where $\vert \alpha^2 \vert \ll 1$, the potential has a point of inflection at~\footnote{The parameters $m_\phi,~A,~\lambda$ are all affected by radiative corrections. However they do not remove the point of inflection nor shift it to unreasonable values. All that happens is that the condition to have a point of inflection~(\ref{dev}) and the Vacuum Expectation Value (VEV) of the inflection point~(\ref{infvev}) will be slightly modified~\cite{AEGJM}.}
\begin{equation}
\phi_0 = \left({m_\phi M^{n-3}_{\rm P}\over \lambda \sqrt{2n-2}}\right)^{1/(n-2)} \, , \label{infvev}
\end{equation}
at which~\cite{AEGJM}
\begin{eqnarray}
\label{pot}
&&V(\phi_0) = \frac{(n-2)^2}{2n(n-1)} m_{\phi}^2\phi_0^2  \, , \\
\label{1st}
&&V'(\phi_0) = \Big({n-2 \over 2}\Big)^2 \alpha^2 m^2_{\phi} \phi_0  \, , \\
\label{2nd}
&&V^{\prime \prime}(\phi_0) = 0 \, , \\
\label{3rd}
&&V^{\prime \prime \prime}(\phi_0) = 2(n-2)^2
{m^2_\phi \over \phi_0} \, .
\end{eqnarray}
We have kept terms that are leading order in $\alpha^2$.

There is a plateau in the vicinity of the point of inflection~\cite{AEGM},
\beq \label{plateau}
\vert \phi - \phi_0 \vert \sim {\phi^3_0 \over 2 n (n-1) M^2_{\rm P}} ,
\eeq
within which the slow roll parameters $\epsilon \equiv (M^2_{\rm P}/2)(V^{\prime}/V)^2$ and $\eta \equiv M^2_{\rm P}(V^{\prime \prime}/V)$ are smaller than $1$. If the field lies in this plateau and has a sufficiently small velocity, inflation occurs~\footnote{Desirable initial conditions for inflation can be naturally set within prior phase(s) of false vacuum inflation either by the help of quantum fluctuations~\cite{AFM}, or as a result of attractor behavior of the inflection point~\cite{ADM3}. Many models of high energy physics possess metastable vacua, and hence can lead to false vacuum inflation at some stage during the evolution of the early universe.}. The Hubble expansion rate during inflation is given by
\begin{equation} \label{hubble}
H_{\rm inf} \simeq \frac{n-2}{\sqrt{6 n (n-1)}}\frac{m_{\phi}\phi_0}{M_{\rm P}}\, ,
\end{equation}
and the total number of e-foldings in the slow-roll regime (where $\vert \epsilon \vert,~\vert \eta \vert < 1$) can be as large as $10^3$, which is more than enough to solve the flatness and isotropy problems.

The number of e-foldings between the time that observationally relevant perturbations exit the horizon and the end of inflation follows~\cite{AEGJM}:
\begin{equation} \label{NCOBE}
{\cal N}_{\rm COBE} \simeq 66.9 + {1 \over 4} {\rm ln} \left[{V(\phi_0) \over M^4_{\rm P}}\right] .
\end{equation}
The amplitude of perturbations thus produced $\delta_H$ and the scalar spectral index $n_s$ are
given by~\cite{Lyth,AEGJM}
\begin{equation} \label{ampl}
\delta_H = {1 \over 5 \pi} \sqrt{{2 \over 3} n(n-1)} (n-2) {m_{\phi} M_{\rm P} \over \phi^2_0}{1 \over \Delta^2}
~ {\rm sin}^2 [{\cal N}_{\rm COBE}\sqrt{\Delta^2}]\, ,
\end{equation}
and
\begin{equation} \label{tilt}
n_s = 1 - 4 \sqrt{\Delta^2} ~ {\rm cot} [{\cal N}_{\rm COBE}\sqrt{\Delta^2}] ,
\end{equation}
where
\begin{equation} \label{Delta}
\Delta^2 \equiv n^2 (n-1)^2 \alpha^2 {\cal N}^{-2}_{\rm COBE} \Big({M_{\rm P} \over \phi_0}\Big)^4\,.
\end{equation}
%

\section{Properties of MSSM inflation}

Experimental data on density perturbations can be used to specify the properties of MSSM inflation. The amplitude of perturbations is~\cite{Liddle} $\delta_H \approx 1.91 \times 10^{-5}$. Then, for weak scale supersymmetry $m_{\phi} \sim 100~{\rm eV}-1$ TeV, Eqs.~(\ref{infvev},\ref{ampl}) require that $n=6$. This singles out inflaton candidates as there are only two flat directions which are lifted by $n=6$ superpotential terms~\cite{GKM}~\footnote{There also exists an inflaton candidate within a minimal extension of MSSM which is represented by the $N H_u L$ flat direction~\cite{AKM} ($N$ and $L$ denote the right-handed sneutrino and left-handed slepton respectively). The inflaton in this case is given by $\phi = ({N} + H_u + {L})/\sqrt{3}$. The simplest extension of the SM gauge group that allows such a flat direction includes $U(1)_{B - L}$, where $B$ and $L$ denote the baryon and lepton number respectively. If neutrinos are Dirac in nature, density perturbations of the correct size will be obtained for neutrino masses of ${\cal O}(0.1~{\rm ev})$~\cite{AKM}, which is the mass indicated by atmospheric neutrino oscillations detected by Super-Kamiokande experiment~\cite{SuperK}.}. One is the $udd$ direction in which case the inflaton is
\begin{equation} \label{udd}
\phi = {{u}^{\alpha}_i + {d}^{\beta}_j + {d}^{\gamma}_k \over \sqrt{3}}.
\end{equation}
Here $u$ and $d$ are the right-handed up- and down-type squarks respectively. The superscripts $1 \leq \alpha,\beta,\gamma \leq 3$ are color indices, and the subscripts $1 \leq i,j,k \leq 3$ denote the quark families. The flatness constraints require that $\alpha \neq \beta \neq \gamma$ and $j \neq k$. The other direction is $LLe$ for which
\begin{equation} \label{LLe}
\phi = {L^a_i + L^b_j + e_k \over \sqrt{3}} \,,
\end{equation}
where $L$ and $e$ are the left-handed and right-handed sleptons respectively. The superscripts $1 \leq a,b \leq 2$ are the weak isospin indices and the subscripts $1 \leq i,j,k \leq 3$ denote the lepton families. The flatness constraints require that $a \neq b$ and $i \neq j \neq k$.

Eqs.~(\ref{hubble},\ref{tilt},\ref{Delta}), combined with the $2 \sigma$ allowed region for the spectral index from 5-year WMAP data $0.934 \leq n_s \leq 0.988$~\cite{WMAP5}, result in
%
\begin{eqnarray}
&&H_{\rm inf} \sim 100~{\rm MeV}-1~{\rm GeV} \, , \label{hubble2} \\
&&2 \times 10^{-6} \leq \Delta^2 \leq 5.2 \times 10^{-6} \, , \label{Delta2}
\end{eqnarray}
and the VEV of the inflection point~(\ref{infvev}) turns out to be
\begin{equation} \label{infvev2}
\phi_0 \sim 10^{14}~{\rm GeV} .
\end{equation}
\begin{figure}
\vspace*{-0.0cm}
\begin{center}
\includegraphics[width=7.0cm]{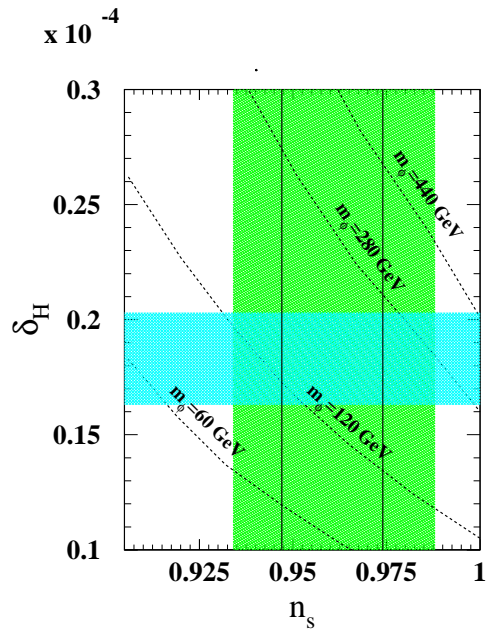}
\caption{$n_s$ is plotted as a function of $\delta_H$ for different
values of $m_{\phi}$. The $2\sigma$ region for $\delta_H$ is shown by the blue horizontal band and the $2\sigma$ allowed region of $n_s$ is shown by the vertical green band. The $1\sigma$ allowed region of $n_s$ is within the solid vertical lines. We choose
$\lambda =1$.} \label{nsdel0}
\end{center}
\end{figure}
%

In figure~1, we show $\delta_H$ as a function of $n_s$ for different values of $m_{\phi}$. The horizontal blue band shows the $2 \sigma$ allowed region for $\delta_H$. The vertical green shaded region is the 2$\sigma$ allowed band for $n_s$, corresponding to Eq.~(\ref{Delta2}), and the region enclosed by solid lines shows the 1$\sigma$ allowed region. This figure is drawn for $\lambda \simeq 1$, see Eq.~(\ref{non}), which is natural in the context of effective field theory~\footnote{Smaller values of $\lambda$ will lead to an increase in $m_{\phi}$~\cite{ADM}.}. It is seen that $m_\phi$ within the $60-440$~GeV range is compatible with the experimentally allowed ranges of $n_s$ and $\delta_H$.
%
%

A remarkable property of MSSM inflation, which is related to inflation occurring near a point of inflection, is that it can give rise to a wide range for scalar spectral index. Indeed it can yield a spectral index within the whole $2 \sigma$ range allowed by 5-year WMAP data $0.934 \leq n_s \leq 0.988$. This stands in contrast with other models (for example, chaotic inflation, hybrid inflation, natural inflation, etc.~\cite{Lyth}) and makes the model very robust~\footnote{Inflection point inflation and sensitivity of its predictions have also been discussed in models of $D$-brane inflation in string theory~\cite{Delicate1,Delicate2}.}.

The low scale of inflation~(\ref{hubble2}) and the sub-Planckian VEV of the inflection point~(\ref{infvev2}) have important consequences:
\begin{itemize}
\item{Gravitational waves that are produced during MSSM inflation are negligible and cannot be detected in future CMB experiments.}
\item{The model is free from the cosmological moduli problem~\cite{Moduli}. The moduli obtain a mass $\sim {\cal O}({\rm TeV})$ from
supersymmetry breaking.
However, since in our case $H_{\rm inf} \sim {\cal O}({\rm GeV})$, quantum fluctuations cannot displace the moduli from
their true minima during the inflationary epoch driven by MSSM flat directions. Moreover, any oscillations of the moduli will be
exponentially damped during the inflationary epoch. This ensures the absence of the cosmological moduli problem in MSSM inflation.}
\item{Supergravity corrections to the inflaton mass are negligible. These corrections typically induce a term $\sim H^2_{\rm inf} \phi^2$ for the
inflaton potential~\cite{DFN,CHRR,GLV,BR}. In our case this is subdominant to the mass term in Eq.~(\ref{scpot}) since $H_{\rm inf} \ll m_\phi$.}
\item{The fact that $\phi_0$ is sub-Planckian guarantees that the inflationary potential is free from the uncertainties about physics at
super-Planckian VEVs. Moreover, the smallness of $H_{\rm inf}$ also precludes any large trans-Planckian correction that would generically go
as $(H_{\rm inf}/M_{\ast})^2$, where $M_{\ast}$ is the scale at which one would expect these effects to show up~\cite{KKLS,BCH}.}
\end{itemize}

\section{MSSM inflation and dark matter}

The inflaton mass $m_{\phi}$ is related to the mass of squarks and sleptons according to
\begin{eqnarray}
&&m^2_\phi = {m^2_{{u}_i} + m^2_{{d}_j} + m^2_{{d}_k} \over 3} ~ ~ ~ ~ ~ (udd ~ {\rm inflaton}) \, , \nonumber \\
&&m^2_\phi = {m^2_{{L}_i} + m^2_{{L}_j} + m^2_{{e}_k} \over 3} ~ ~ ~ ~ ~ (LLe ~ {\rm inflaton}) \, . \nonumber
\end{eqnarray}
The bound on $m_{\phi}$ from density perturbations will then be translated into the bounds on the scalar masses. These bounds apply to the masses at a scale $\sim \phi_0$, see Eq.~(\ref{infvev}), around which inflation occurs. To make a connection with sparticle masses at the weak scale, which will be probed at colliders, one should use appropriate Renormalization Group Equations (RGEs). The one-loop RGEs are given by~\cite{AEGJM,ADM}
\begin{eqnarray}
\mu{dm_{\phi}^2 \over {d\mu}} &=& -{1 \over {6\pi^2}}({4} {M_3^2} g_3^2+ {2 \over {5}}{M_1^2} g_1^2) ~ ~ ~ ~ ~ (udd ~ {\rm inflaton}) \, , \nonumber \\
\mu{dm_{\phi}^2 \over {d\mu}} &=& -{1 \over {6\pi^2}}({3 \over 2}{M_2^2} g_2^2 + {9 \over {10}} {M_1^2} g_1^2) ~ ~ ~ ~ ~ (LLe ~ {\rm inflaton}) \, .
\end{eqnarray}
Here $g_1,~g_2,~g_3$ and $M_1,~M_{2},~M_3$ are gauge couplings and gaugino masses of $U(1),~SU(2),~SU(3)$ respectively. Explicit calculations can be done in the case of minimal supergravity (mSUGRA), which is motivated by unification of gauge couplings at the Grand Unified Theory (GUT) scale $M_{\rm G} \simeq 2 \times 10^{16}$ GeV.  The models of mSUGRA depend only on four parameters and one sign: $m_0$ (the universal scalar mass at $M_{\rm G}$); $m_{1/2}$ (the universal gaugino mass at $M_{\rm G}$); $A_0$ (the universal trilinear $A$-term at $M_{\rm G}$); $\tan\beta = \langle H_u \rangle / \langle H_d \rangle$ at the electroweak scale (where $H_u$ and $H_d$ give masses to up-type and down-type quarks respectively); and the sign of $\mu$, the Higgs mixing parameter in the superpotential ($W_{\mu} = \mu H_u H_d$).
%
\begin{figure} [t]
\vspace{1cm} \center
\includegraphics[width=8.5cm]{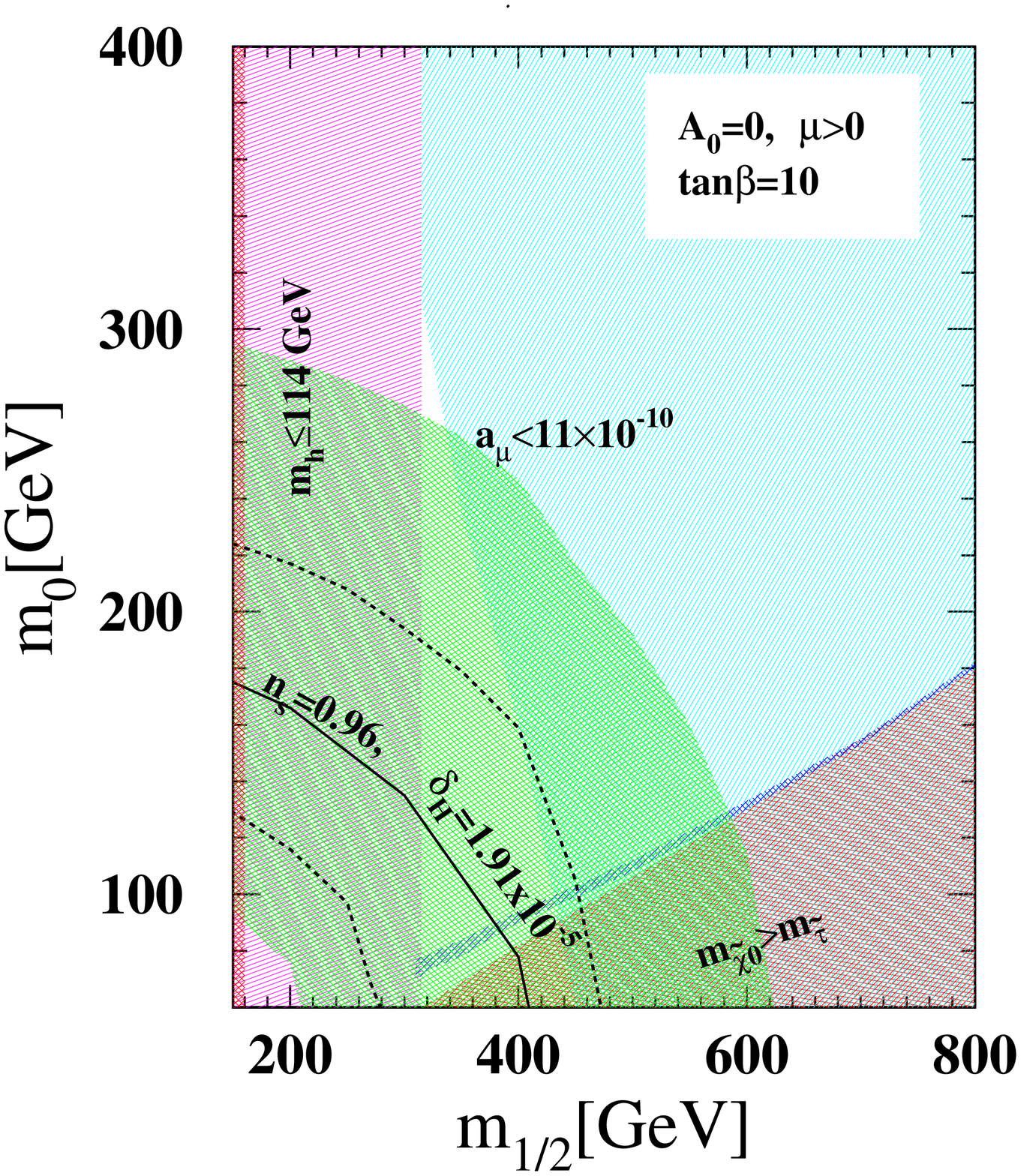}
\includegraphics[width=8.5cm]{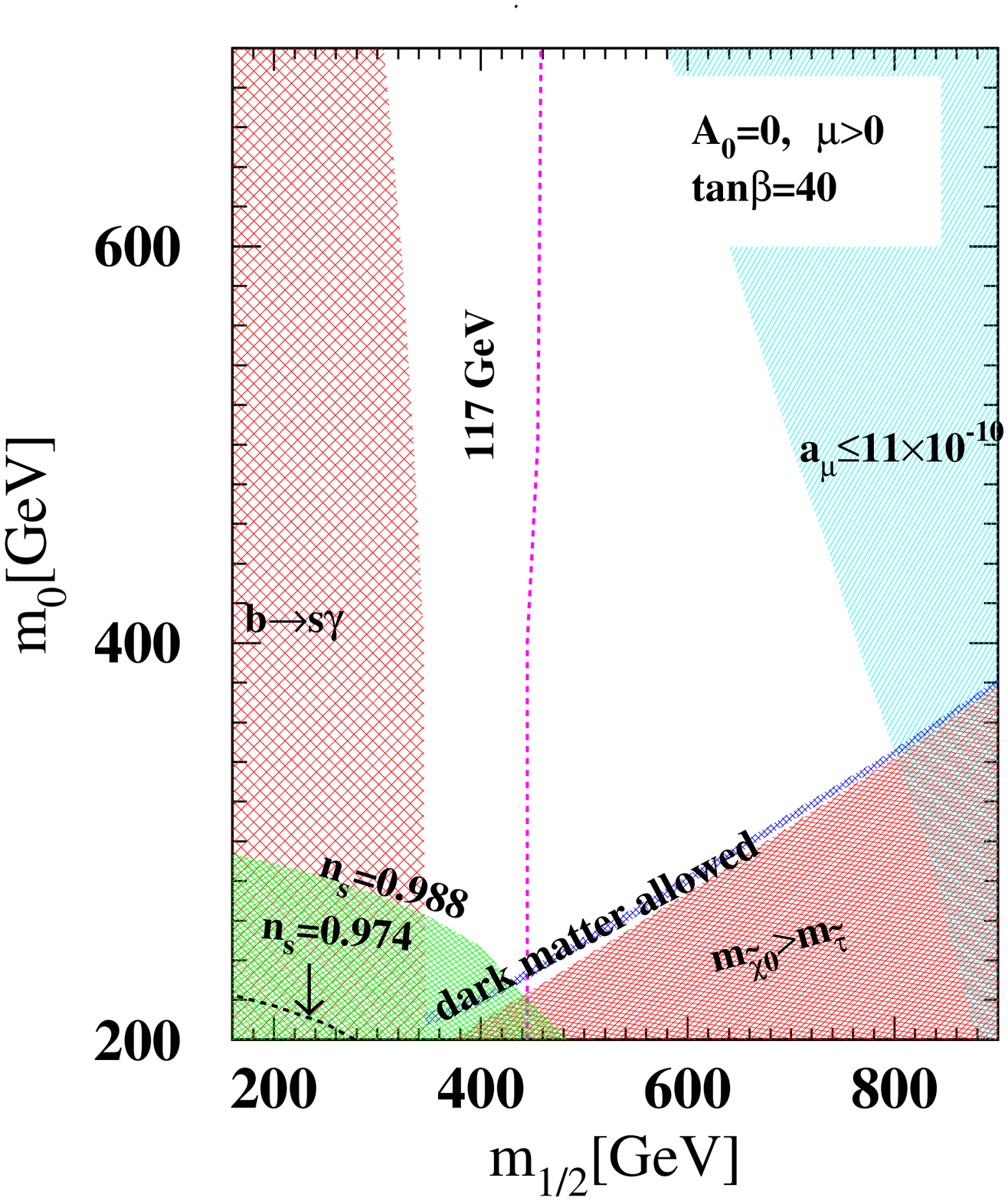}
\caption{The contours for 2$\sigma$ allowed values of $n_s$ are shown in the $m_0-m_{1/2}$
plane by the green band for ${\rm tan}\beta = 10$ (left panel) and ${\rm tan}\beta =40$ (right panel). The 1$\sigma$ contours are shown by the dotted lines. The dark matter allowed regions {narrow blue
corridors} correspond to the stau-neutralino co-annihilation and the focus point regions respectively. We show the (g-2)$_\mu$ region (light blue) for $a_{\mu}\leq 11\times10^{-8}$, Higgs mass $\leq 114$ GeV (pink region) and LEPII
bounds on the mass of supersymmetric particles (red). The black region is not allowed by radiative electroweak symmetry breaking.}
\end{figure}
%

The model parameters are already significantly constrained by different experimental results~\cite{PDG,Muon,LEP,CLEO}. A further constraint arises from the he requirement that the Lightest Supersymmetric Particle (LSP), which is stable in models with conserved $R$-parity, has the right relic abundance to be the cold dark matter (CDM) in the universe. The $1 \sigma$ bound from latest cosmological data~\cite{WMAP5} gives a relic density bound of $0.220 < \Omega_{\rm CDM} < 0246$ for CDM. In mSUGRA the lightest neutralino is the dark matter candidate.
%
%
%
%
%
%
%
%
The allowed parameter space, at present, has mostly three distinct regions selected out by dark matter constraints~\cite{EOSS,ADH,BBBK,LN,CCN,BG,DDK,FMW}: (i) the stau-neutralino coannihilation region, (ii) the focus point region, and (iii) the scalar Higgs annihilation funnel region.
%
%
%
%
%
%

We show the mSUGRA parameter space in figure~2 for $\tan \beta=10,~40$ with the $udd$ flat direction as the inflaton using $\lambda \simeq 1$ (the figure for the $LLe$ flat direction is similar)~\footnote{Inflation and dark matter can be unified in the case that the $N H_u L$ flat direction is the inflaton~\cite{ADM2}. The inflaton $\phi = ({N} + H_u + {L})/\sqrt{3}$ has a right-handed sneutrino component which can be the dark matter candidate. Sneutrino dark matter can be seen in the upcoming direct detection experiments and also be produced at the LHC. For more details, see Ref.~\cite{ADM2}.}. The contours correspond to different values of $n_s$ within the $2 \sigma$ range allowed by 5-year WMAP data,
for $\delta_H = 1.91 \times 10^{-5}$. The constraints on the parameter space arising from inflation are compatible with those from dark matter and other experimental results. It is seen that $\tan \beta$ needs to be smaller to allow for smaller values of $n_s$. It is also interesting to note that the allowed region of $m_{\phi}$ lies in the stau-neutralino coannihilation region which requires smaller values of the supersymmetric particle masses. The supersymmetric particles in this parameter space are, therefore, within the reach of the LHC very quickly. The
detection of this region at the LHC has been considered in Ref.~\cite{Bhaskar}.
\section{Reheating after MSSM inflation}

After the end of inflation, the inflaton~(\ref{udd},\ref{LLe}) rolls towards
the global minimum of its potential at the origin. At this stage the dominant term in the scalar
potential will be $m_\phi \phi^2/2$, see Eq.~(\ref{scpot}), which results in oscillations with a frequency of $m_\phi$. Since $m_{\phi} \sim 10^3 H_{\rm inf}$, the inflaton oscillates about the origin a large number of times within the first Hubble time after the end of inflation. Hence the effect of expansion is negligible.
%
%
%
Since the inflaton is a linear combination of squarks~(\ref{udd}) or sleptons~(\ref{LLe}), it has gauge couplings to the
gauge/gaugino fields and Yukawa couplings to the Higgs/Higgsino
fields.
To elucidate the physics, we consider the case when the ${\bf LLe}$ flat direction is the
inflaton. The situation for the $udd$ flat direction as the inflaton is similar.




The VEV of the inflaton spontaneously breaks $SU(2)\times U(1)$ symmetry, and therefore, induces a supersymmetry
conserving mass to the electroweak gauge/gaugino
fields (similar to what happens in electroweak symmetry breaking fashion via the Higgs mechanism). When the flat direction goes to its minimum this mass vanishes, and the gauge symmetry is restored. However, the mass undergoes a non-adiabatic time variation every time that the inflaton crosses the origin. This results in an efficient creation of gauge and gaugino quanta with a
physical momentum $k \ls \left(g m_{\phi} \phi_0 \right)^{1/2}$ within a short interval $\Delta t \sim \left(g m_{\phi} \phi_0
\right)^{-1/2}$, where $\phi_0$ is given by Eq.~(\ref{infvev}) and $g$ is the corresponding gauge coupling. The number density of the gauge/gaugino quanta thus produced is given by~\cite{KLS1,KLS2}
\begin{equation} \label{chiden}
n_{g} \approx {\left(g m_{\phi} \phi_0
\right)^{3/2} \over 8 \pi^3}\,.
\end{equation}
As the inflaton VEV is rolling back to its maximum value $\phi_0$, the
mass of the produced quanta increases. The
gauge and gaugino fields can (perturbatively) decay to the fields
which are not coupled to the inflaton, for instance to (s)quarks. Note
that (s)quarks are not coupled to the flat direction, hence they
remain massless throughout the oscillations. The total decay rate of
the gauge/gaugino fields is then given by $\Gamma = C \left(g^2/48\pi
\right) g\phi $, where $C \sim {\cal O}(10)$ is a numerical factor
counting the multiplicity of final states.

The decay of gauge and gaugino quanta happens very quickly and converts a fraction $f$ of the inflaton energy density (for details, see~\cite{AEGJM}) to relativistic (s)quarks where
%
%
%
%
%
\begin{equation} \label{ratio}
f \sim 10^{-2} g\, .
\end{equation}
%
This is the so-called instant preheating mechanism~\cite{INSTANT}.
The rapid conversion of fraction $f$ of the inflaton energy density into relativistic particles happens twice in each oscillation. Note that there will be hundreds of oscillations within the first few Hubble times after the end of inflation since $m_\phi \sim 10^3 H_{\rm inf}$.
Reheating is therefore quite efficient in this model as almost all the energy density in the
inflaton will decay into radiation within a couple of Hubble times. The resulting reheat temperature of the universe will be
\begin{equation} \label{reheat}
T_{\rm R} \sim 10^7~{\rm GeV} ,
\end{equation}
which is sufficiently low to avoid thermal overproduction of gravitinos (for more details, see~\cite{AEGJM}).

\section{Discussion and Conclusion}

The existence of a point of inflection in the scalar potential of two MSSM flat directions provides all the necessary ingredients for a realistic and successful model of inflation.
The exceptional feature of the model, which sets it apart from conventional singlet field inflation models, is that here the inflaton is not added as an ad hoc field whose sole purpose is to drive inflation. Instead it is a combination of the squark and slepton fields, and hence its couplings to the matter and gauge fields are known. This not only gives the inflaton a natural place within particle physics, but also makes it possible to address reheating after inflation in an unambiguous way.



MSSM inflation occurs at a low scale, corresponding to a Hubble expansion rate $H_{\rm inf} \sim {\cal O}({\rm GeV})$ and sub-Planckian field values. This implies negligible supergravity and trans-Planckian corrections and solves the cosmological moduli problem. The model is robust as it can give rise to density perturbations of the correct size with a scalar spectral index in the entire range allowed by the 5-year WMAP data. The parameter space for inflation is compatible with that of supersymmetric dark matter.
%
$\lambda \sim {\cal O}(1)$ (as expected in an effective field theory approach) can be explained.
In the context of mSUGRA
the stau-neutralino coannihilation region is most preferred to satisfy the dark matter content of the universe. The masses of supersymmetric particles in this region are mostly within the reach of the LHC.

Reheating after MSSM inflation is very efficient. Non-perturbative production of gauge and gaugino fields and their subsequent decay to relativistic particles result in a reheat temperature $T_{\rm R} \sim 10^7$ GeV despite the small value of $H_{\rm inf}$. This is sufficiently high for various mechanisms of generating the baryon asymmetry of the universe and produce thermal dark matter while avoiding overproduction of gravitinos.

The existence of the inflection point requires a fine-tuning of the ratio of the soft supersymmetry breaking parameters $m_\phi$ and $A$.
Radiative corrections change $A/m_{\phi}$ only slightly so that inflection point inflation can always be achieved for some value of this ratio.
A more detailed investigation is required to address the fine-tuning issue, but it is warranted by the success of MSSM inflation, which is unique in being both a successful model of inflation and at the same time having a concrete and real connection to physics that can be observed in earth-bound laboratories.

\section{Acknowledgments}

The author is indebted to Bhaskar Dutta, Kari Enqvist, Andrew Frey, Juan Garcia-Bellido, Asko Jokinen, Alex Kusenko and Anupam Mazumdar for collaboration and numerous discussions on various aspects of the physics presented here. He wishes to thank Robert Brandenberger, Cliff Burgess, Manuel Drees, Gordy Kane, Justin Khoury, David Lyth, Guy Moore, Scott Thomas and Maxim Pospelov for valuable discussions on this subject. Katie Richardson-McDaniel is acknowledged for careful reading of this manuscript and useful suggestions. This research was supported in part by Perimeter Institute for Theoretical Physics.



\end{document}